\documentclass[12pt,twoside]{article}   
\usepackage{latexsym} 
\usepackage{mathptm}
\usepackage{amsmath}
\usepackage{amssymb}
\usepackage{graphicx}

\begin{document}
 
\title{Gravity can be neither classical nor quantized}
\author{Sabine Hossenfelder \thanks{hossi@nordita.org}\\
{\footnotesize{\sl Nordita, Roslagstullsbacken 23, 106 91 Stockholm, Sweden}}}
\date{}
\maketitle
 
\begin{abstract}
I argue that it is possible for a theory to be neither quantized nor classical. We should
therefore give up the assumption that the fundamental theory which describes gravity at 
shortest distances must
either be quantized, or quantization must emerge from a fundamentally  classical theory.
To illustrate my point I will discuss an example for a theory that is neither classical nor quantized, and
argue that it has the potential to resolve the tensions between the quantum field
theories of the standard model and general relativity.
\end{abstract}

\section*{To quantize or not to quantize gravity}

Gravity stands apart from the other three interactions of the standard model by its refusal
to be quantized. To be more precise, quantizing gravity is not the actual problem;
gravity can be perturbatively quantized. The problem is that the so quantized theory cannot
be used at energies close by and above the Planck energy, and thus cannot be considered a fundamental
theory; it is said to be `non-renormalizable,' meaning it has no predictive power in
the extremely high energy regime. 

This mismatch between the quantum field theories of the standard model and classical
general relativity is more than an aesthetic problem: It signifies a severe shortcoming
of our understanding of nature. This shortcoming has drawn a lot of attention because
its resolution it is an opportunity to completely overhaul our understanding of space, time
and matter. The search for a consistent theory of quantum gravity that could be applied
also at Planckian energies, or strong curvature respectively, has thus lead to many 
proposals. But progress has been slow and in the absence of experimental evidence, 
our reasons for the necessity of quantizing gravity are theoretical:

\begin{enumerate}
\item Classical general relativity
predicts the formation of singularities under quite general circumstances. 
Such singularities are unphysical and should not occur in a fundamentally meaningful theory. It is
expected that quantum gravity is necessary to prevent the formation of singularities.

\item Applying quantum
field theory in a curved background at small curvature leads to the evaporation of 
black holes, as first shown by Hawking \cite{Hawking}. This black hole evaporation
however seems to violate unitary which is incompatible with quantum mechanics. 
It is widely believed that quantum gravitational effects restore unitarity and
information is conserved. 

\item All quantum fields carry energy so they all need to
couple to the gravitational field, but we do not know a consistent way to couple a quantum
field to a classical field. As Hannah and Eppley have argued \cite{EH}, the attempt to do such a 
coupling leads either to a violation of the uncertainty principle (and thus would necessitate a 
change of the quantum theory) or to the possibility of superluminal signaling, which brings more
problems than it solves. While Mattingly has argued \cite{Mattingly} that Hannah and Eppley's thought experiment
can not be carried out in our universe, that does not address the problem of consistency.
\end{enumerate}

These issues have all been extensively studied and discussed in the literature and
are familiar ground. The
most obvious way to address them seems to be a non-perturbative theory in one or
other form, and several attempts to construct one are under way. I will use the
opportunity of the essay contest to stray from the well-trodden ground and argue that we should instead
reinvestigate the apparent tension between the quantized matter and non-quantized
gravity. It is worthwhile for the following to recall the problems with coupling a classical to
a quantum field. 

The first problem, as illuminated by Hannah and Eppley is that the classical and the
quantum fields
would have different uncertainty relations, and their coupling would require
a modification of the quantum theory. Just coupling them as they are leads to an inconsistent
theory. The beauty of Hannah and Eppley's tought argument is its generality, but that
is also its shortcoming, because it does not tell us how a suitable modification of
quantum theory could allow such a coupling to be consistent.

The second problem is that it is unclear how mathematically the coupling should be realized,
as the quantum field is operator-valued and the classical field is a function on space-time. 
One possible answer to this is that any function can be identified with an operator on
the Hilbert space by multiplying it with the identity. However, the associated operators 
would always be commuting, so they are of limited use to construct a geometrical quantity
that can be set equal to the operator of the stress-energy-tensor ({\sc SET}) of the quantum fields.

Another way to realize the coupling is to construct  classical field from the
operator of the {\sc SET} by taking the expecation value. The problem
with this approach is that the expectation value may differ before and after measurement,
which then conflicts with the local conservation laws of general relativity. Coupling the
classical field to the {\sc SET}'s expectation value is thus usually
considered valid only in approximation when superpositions carry negligible amounts of
energy. 

Because of these difficultites to make sense of the theory, leaving gravity classical 
while the other interactions are quantized is not a very promising option. However,
this theoretical assessment should be supported by experimental test; recent proposals 
for this have been put forward in \cite{Giulini:2011uw,vanMeter:2011xr}.

\section*{How to be neither classical nor quantized}

Let us carefully retrace the logic of the arguments in the previous section. 

 We have experimental evidence that matter is quantized
in the energy regimes that we have tested. We cannot leave gravity unquantized if it
couples to quantized matter. Thus gravity has to be quantized in the energy
regimes we have tested. We can quantize gravity perturbatively. This theory
does make sense in the energy regimes that we have tested, but does not make sense in the
strong curvature regime. We have no experimental
evidence for the existence and properties of singularities or black hole evaporation, or the
behavior of matter in the strong curvature regime. 

To conclude from the previous paragraph that we need a non-perturbative completion of quantum gravity
necessitates a further assumption, that is that the quantization procedure itself 
is independent of the energy range at which we apply the theory. It is this
assumption that I argue should be given up. 

We 
normally think of a theory as either being quantized or classical, but let us
entertain the
possibility that quantization is energy-dependent. Concretely, consider that
Planck's constant $\hbar$ is a field whose value at high energies goes to zero. In four 
space-time dimensions, 
Newton's constant is $G= \hbar c/m_{\rm Pl}^2$, so if we keep mass units fix, $G$ will go 
to zero together with $\hbar$, thereby decoupling gravity. If gravity decouples, there's
no reason for singularities to form. If gravity becomes classical, there's no problem with
the perturbative expansion. So this possibility seems intriguing, if somewhat vague.
I will now make this idea more concrete and then explain how it addresses
the previously listed problems with quantizing gravity.

The starting point is that Planck's constant is a massless scalar field over
space time $\hbar(x,t)$, and the equal time commutation relations for all fields, including
Planck's constant itself, are proportional then to $\hbar(x,t)$. Since we have no experimental 
evidence for the variation of Planck's constant, the
most conservative assumption is that the $\hbar$-field is presently in its
ground state, and difficult to excite with energies that we have access to. This
suggests that we think about quantization as the consequence of a spontaneous
symmetry breaking, and we have to add a suitable potential for $\hbar$ to the Lagrangian to achieve 
this. We are presently experiencing $\hbar(x,t)$ as having a non-zero
vacuum expectation value that we will denote with $\hbar_0$. This is the measured
value of Planck's constant. But at high temperature, presumably close by the Planck
energy, the symmetry can be restored,
resulting in a classical theory.

Gravity and matter then
have a quantized phase and an unquantized phase, and are fundamentally neither
quantized nor classical in the same sense that water is 
fundamentally neither liquid nor solid. Quantization, in this case, is also not
emergent from a classical theory because the condition for second
quantization does always contain the $\hbar(x,t)$. 

\section*{A new look at old problems}

Let us now come back to the three problems mentioned in
the first section that
a theory for quantum gravity should address. 

First, there is the
formation of singularities. We know of two types of singularities
that we should worry about, the Big Bang singularity and the
singularities inside black holes. 

If we move backwards in time towards the early universe, the
temperature of matter increases and will eventually exceed the
Planck energy. This is the standard scenario in which symmetry
restoration takes place \cite{Kapusta}, so the expectation value of $\hbar$ goes
to zero, gravity becomes classical, and matter decouples. If
matter decouples, it cannot collapse to a singularity. 

Collapse to a black hole is somewhat more complicated because
it's not a priori clear that the temperature of the collapsing
matter necessarily increases, but it plausibly does so for the
following reason\footnote{I acknowledge helpful conversation
with Cole Miller on this issue.}. If matter collapses to a black hole, it 
does so rapidly and after horizon
formation lightcones topple inward, so no heat exchange with 
the environment can take place and
the process is adiabatic. The entropy of the degenerate Fermi 
gas is proportional to $T n^{-2/3}$, where $T$ is the temperature
and $n$ is the number density. This means that if the number density rises and entropy 
remains constant, the temperature has to rise 
\cite{Kothari}. So again, matter decouples and there is nothing
left to drive the formation of singularities. 

Note that the $\hbar$-field makes a contribution to the source term, necessary for energy 
conservation.

Second, there is the black hole information loss. It was
argued in \cite{Hossenfelder:2009xq} that the problem is
caused by the singularity, not the black hole horizon, and
that removing the singularity can resolve the information
loss problem. This necessitates the weak interpretation of the 
Bekenstein-Hawking entropy so that a stable or quasi-stable Planck
scale remnant, or a baby-universe, can store a large amount of information. There
are some objections to the existence of such remnants, but
they rely on the use of effective field theory in strong
curvature regimes, the validity of which is questionable \cite{Donoghue:2009mn}.
Thus, unitarity in black hole evaporation can be addressed by
the first point, avoiding the formation of singularities.

Third, the difficulty of coupling a quantum field to a classical
field and the non-renormalizability of perturbatively
quantized gravitty. In the here proposed scenario, there is never 
a classical field coupled to a quantum
field. Instead, gravity and matter are of the same type and together
either in a quantum phase or a classical phase. In the quantum
phase, gravity is quantized perturbatively. It then needs to
be shown that the perturbation series cleanly converges for
high energy scattering because $\hbar$ is no longer a constant. 
This is a subtle point and I can here only
give a rough argument.

To see how this would work, first note that we can rewrite the
equal time commutation relation into a commutation relation for
annihilation and creation operators of the fields. The commutator
between annihilation and creation operators is then proportional
to the Fourier-transform of $\hbar(x,t)$, which I will denote $\tilde \hbar$. 
The same is true for the annihilation and creation operators of
$\hbar(x,t)$ (though the prefactors differ for dimensional reasons).

Now consider an arbitrary $S$-matrix 
transition amplitude with some interaction vertices. We evaluate it by
using the commutation relations repeatedly until annihilation operators are shifted to the very
right side, acting on the vacuum, which leaves $c$-numbers, or the Feynman rules respectively. 
If Planck's constant is a field, then
every time we use the commutation relation, we get
a power of the $\hbar$-field, and the $S$-matrix expansion is a series in 
expectation value of powers 
of $\tilde \hbar$ times the other factors of the transition amplitudes. Then,
we use the commutation relations on $\hbar$, or its annihilation and
creation operators respetively. Now note that exchanging two of these will
only give back one $\tilde \hbar$. Thus, we can get rid of the expectation 
value of powers, so that in the end we will have a
series in powers of vacuum expectation values of $\tilde \hbar$ (as opposed
to a series of expectation values of powers, note the difference).

If we
consider the symmetry breaking potential to be induced by quantum corrections at low order,
the transition to full symmetry restoration may be at a finite value of energy. In this case
then, the quantum corrections which would normally diverge would cleanly go to zero, removing
this last problem with the perturbative quantization of gravity.

\section*{Summary}

I have argued that the fundamental theory can be neither classical nor
quantized, but that quantization may be a phase that results from 
spontaneous symmetry breaking. Needless to say, this proposal is presently
very speculative and immature. Some more details can be found in
\cite{own}, but open questions remain. However, I hope to have
convinced the reader that giving up the assumption that a theory
is either classical or quantized can be fruitful and offers a
new possibility to address the problems with quantum gravity.

\end{document}